\documentclass[mathleft
]{an}
\usepackage{graphicx}
\usepackage{times}
\begin{document}

\Pagespan{999}{}
\Yearpublication{2010}%
\Yearsubmission{2009}%
\Month{1}%
\Volume{999}%
\Issue{99}%

\title{High-fidelity spectroscopy at the highest resolutions}

\author{Dainis Dravins\thanks{Corresponding author:
  \email{dainis@astro.lu.se}\newline}
}
\titlerunning{High-fidelity spectroscopy}
\authorrunning{D.Dravins}
\institute{Lund Observatory, Box 43, SE-22100 Lund, Sweden}

\received{1 January 2010}
\accepted{1 January 2010}
\publonline{later}


\abstract{High-fidelity spectroscopy presents challenges for both observations and in designing instruments. High-resolution and high-accuracy spectra are required for verifying hydrodynamic stellar atmospheres and for resolving intergalactic absorption-line structures in quasars.  Even with great photon fluxes from large telescopes with matching spectrometers, precise measurements of line profiles and wavelength positions encounter various physical, observational, and instrumental limits.  The analysis may be limited by astrophysical and telluric blends, lack of suitable lines, imprecise laboratory wavelengths, or instrumental imperfections.  To some extent, such limits can be pushed by forming averages over many similar spectral lines, thus averaging away small random blends and wavelength errors. In situations where theoretical predictions of lineshapes and shifts can be accurately made (e.g., hydrodynamic models of solar-type stars), the consistency between noisy observations and theoretical predictions may be verified; however this is not feasible for, e.g., the complex of intergalactic metal lines in spectra of distant quasars, where the primary data must come from observations.  To more fully resolve lineshapes and interpret wavelength shifts in stars and quasars alike, spectral resolutions on order R=300,000 or more are required; a level that is becoming (but is not yet) available.  A grand challenge remains to design efficient spectrometers with resolutions approaching R=1,000,000 for the forthcoming generation of extremely large telescopes. }

\maketitle

\section{Why the highest resolution?}

Sophisticated astrophysical models often predict spectral-line shapes with asymmetries and wavelength shifts (e.g., 3-dimensional hydrodynamics in stellar atmospheres; isotopic constituents of interstellar lines; velocity gradients in regions of line formation; or simply the hyperfine structure of the atomic transitions themselves).  However, the confrontation with observations is often limited by blends, lack of suitable lines, imprecise laboratory wavelengths, insufficient spectral resolution and instrumental imperfections.  Observational limits can be pushed by averaging many similar lines, thus averaging small random blends and wavelength errors, although the resolution cannot be increased beyond that at which the original data are sampled.  Instrumental designs are often limited by the optical interface of high-resolution spectrometers to [very] large telescopes with their [very] large image scales.

\section{Solar and stellar spectra}

Naturally, high-resolution spectroscopy was first explored for the brightest sources such as the Sun and brighter stars.  This has also enabled paradigm changes in the analysis of stellar atmospheres in that, contrary to the days of classical one-dimensional homogeneous and static models, it has been realized that it is not possible -- not even in principle -- to infer detailed stellar properties from analyzing observed line parameters alone, no matter how precisely the spectrum would be measured.  Any stellar absorption line is built up by great many contributions from a wide variety of temporally variable inhomogeneities across the stellar surface, whose statistical averaging over time and space produce the line shapes and shifts that can be observed in integrated starlight. While it is possible to compute resulting line profiles (with their asymmetries and wavelength shifts) from hydrodynamic models, the opposite is not feasible because the 3-dimensional structure cannot be uniquely deduced from observed line shapes alone (e.g., Asplund 2005).  

While hydrodynamic models may predict detailed properties for a hypothetical spectral line, a confrontation with observations may be unfeasible because stellar lines of the desired species, strength, ionization level, etc., might simply not exist, or be unobservable in practice.  In real spectra, lines are frequently blended by stellar rotation, by overlapping telluric lines from the terrestrial atmosphere or else smeared by inadequate instrumental resolution, in practice precluding more detailed studies. 

While, in the past, it may have been sufficient to merely resolve the presence of lines, and to determine their strengths, finding line asymmetries implies measurements over smaller fractions of each line-width.  Any point on a spectral-line bisector (median) is obtained from intensities at two wavelength positions on either side of the line center.  To define not only the bisector slope, but also its curvature requires at least some five points, implying at least ten measurements across the line profile.  Given a width of a typical photospheric line of, say, 12~km~s$^{-1}$, an ordinary resolution of R~=~100,000 (3~km~s$^{-1}$) can only indicate the general sense of line asymmetry.  Both simulated and observed spectra show how the bisectors degrade when the spectral resolutions decrease towards such values (e.g., Dravins \& Nordlund 1990b; Allende Prieto et al.\ 2002; Ram{\'\i}rez et al.\ 2009).  The general appearance of line profiles recorded under different resolutions is shown in Fig.~1.

\begin{figure}
\includegraphics[width=8.3cm]{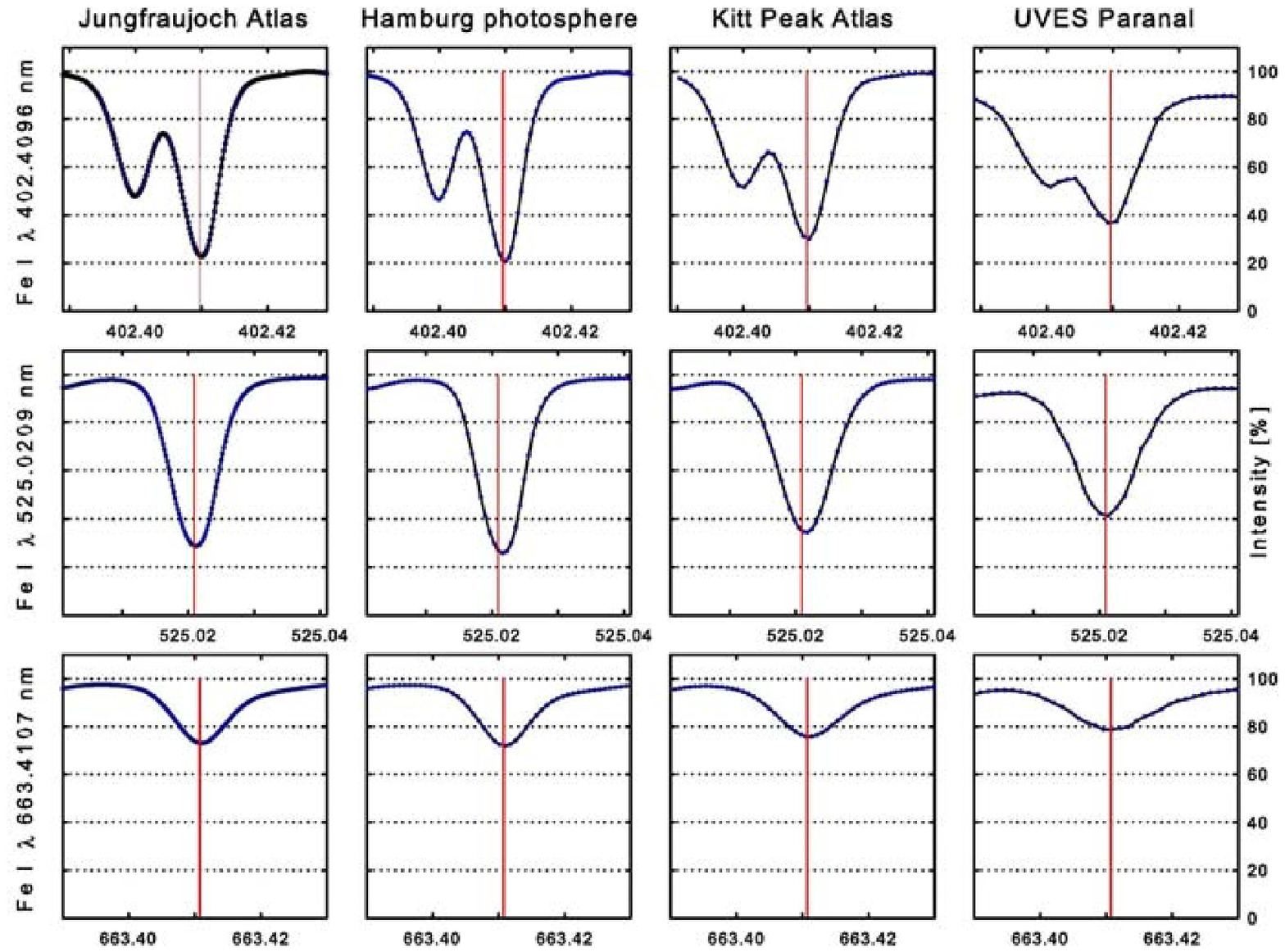} 
\caption{Individual line profiles in solar spectra recorded at different resolutions.  Three representative \ion{Fe}{i} lines are shown: a deep but somewhat blended line in the violet, a clean line in the green, and a weaker line in the red.  Left to right: spectra of solar disk center recorded with a grating spectrometer at R~=~$\lambda$/$\Delta\lambda \approx$~700,000 (Delbouille et al.\ 1989); solar disk center (Neckel 1999) and integrated sunlight (Kurucz et al.\ 1984), both with a Fourier-transform spectrometer at R~$\approx$~350,000; and moonlight with a stellar spectrometer at R~$\approx$~80,000 (Bagnulo et al.\ 2003).  Vertical lines mark laboratory wavelengths and thus the `naively' expected line positions. (Dravins 2008) }
\label{Fig1}
\end{figure}

\begin{figure}
\includegraphics[width=8.1cm]{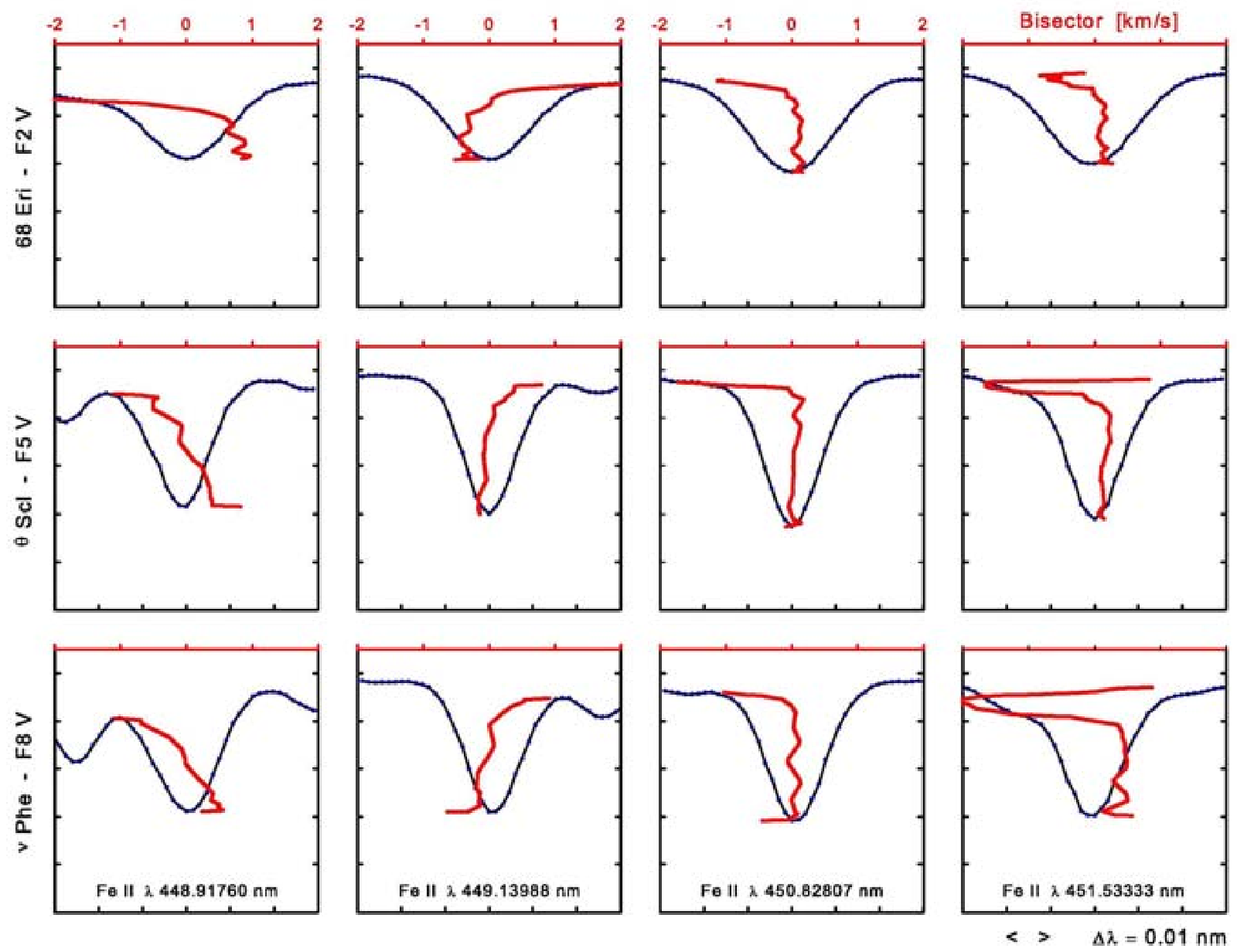} 
\caption{Examples of individual line bisectors (medians), overplotted on the absorption-line profiles, for three representative \ion{Fe}{ii} lines in R~$\approx$~80,000 spectra (Bagnulo et al.\ 2003) of different F-type stars: 68~Eri (F2~V), $\theta$~Scl (F5~V), and $\nu$~Phe (F8~V).  The bisector scale (top) is expanded a factor of 10 relative to the wavelength one. (Dravins 2008) }
\label{Fig2}
\end{figure}

\section{Advancing high-resolution spectroscopy} 

High spectral resolution alone may not suffice to obtain high-fidelity spectra.  The desired line may be blended with stellar or telluric ones; its laboratory wavelength could be uncertain; the data may be noisy, and the line might be distorted due to stellar rotation or oscillations.

Fig.~2 illustrates the difficulty of finding truly `unblended' lines.  Visually, all lines selected appear quite clean but comparing the same line between different stars, it is seen that while bisectors often share some common features, the bisector shapes differ strongly among different lines in the same star (where, under similar conditions of line formation, one would expect them to be similar).  Thus, the `noise' -- the cause of bisector deviations from a representative mean -- does not originate from photometric errors but instead is largely `astrophysical' in character, caused by blending lines and similar (Dravins 2008).  These spectra represent current high-fidelity spectroscopy, taken from the UVES Paranal Project (Bagnulo et al.\ 2003), where data of particularly low noise were recorded at resolution R~$\approx$~80,000 with the UVES spectrometer (Dekker et al.\ 2000) at the ESO VLT {\it{Kueyen}} unit.

\subsection{Averages over similar lines}

The omnipresence of weak blends makes it impossible to extract reliable asymmetries or shifts from any single line only. However, effects of at least slight blends may be circumvented by forming averages over groups of similar lines which can be expected to have similar physical signatures.  Fig.~3 shows such average \ion{Fe}{ii} bisectors for both solar disk center (Delbouille et al.\ 1989) and for integrated sunlight (Kurucz et al.\ 1984).  The wavelength scale now is absolute, using laboratory data from the \textsc{FERRUM} project (Johansson 2002; Dravins 2008).  For solar disk center, 104 lines were found to be clean enough for such averaging, and 93 for integrated sunlight.  From such numbers, reasonably well-defined mean bisectors emerge; however, most atomic species have rather fewer lines.

\begin{figure}[]
\centering
\includegraphics[width=6cm]{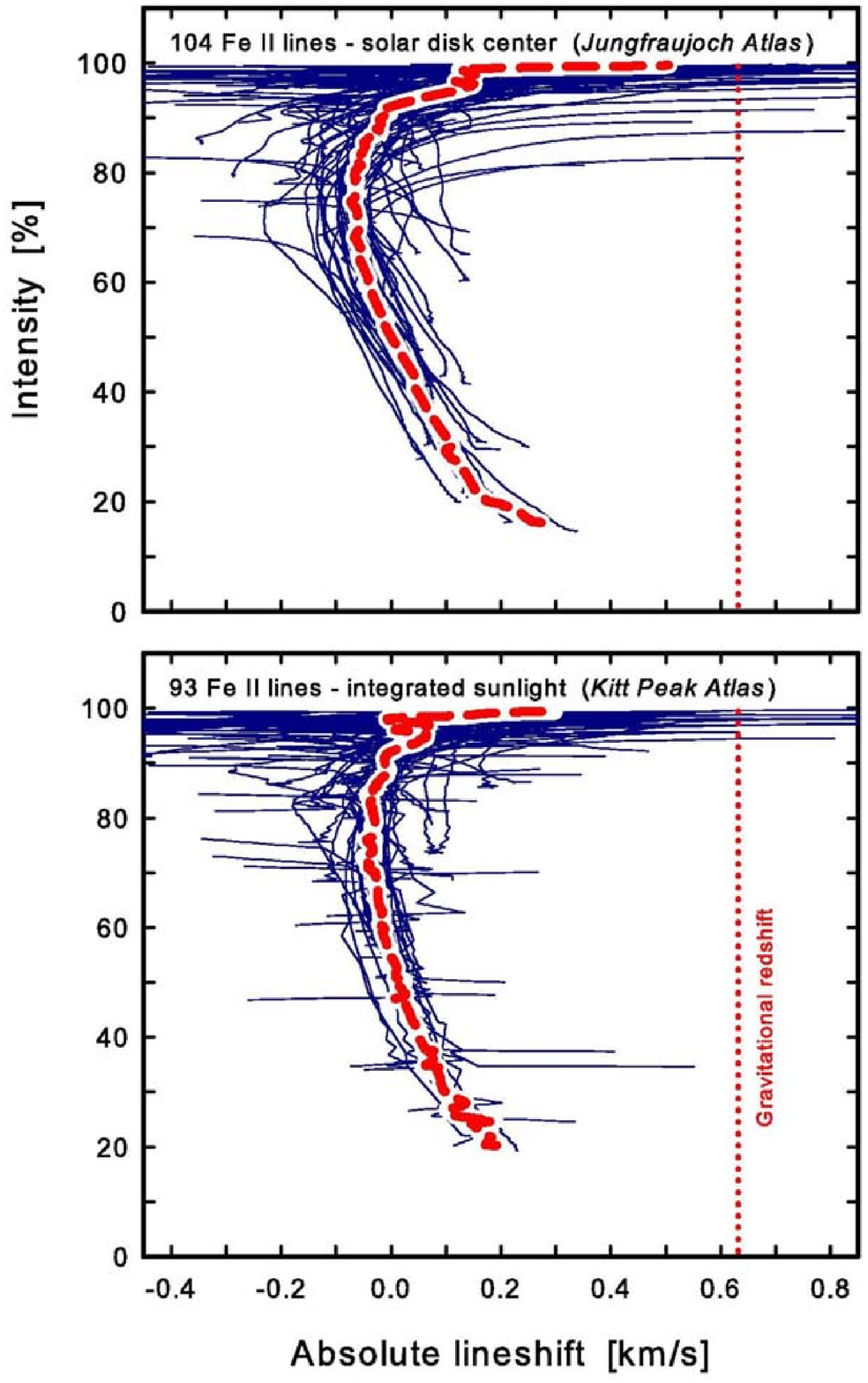} 
\caption{ \ion{Fe}{ii} bisectors in spectra of solar-disk center and of integrated sunlight.  Each thin curve is the bisector of one spectral line; the thick dashed is their average.  Such averaging over many similar lines (which can physically be expected to have similar intrinsic shapes) is required to reduce effects of weak random blends.  The smaller bisector curvature for integrated sunlight reflects both intrinsic line-profile changes across the solar disk, and line broadening due to solar rotation.  The vertical scale denotes the intensity in units of the spectral continuum, while the dotted line marks the wavelength expected in a classical 1-dimensional model atmosphere, given by the laboratory wavelength, displaced by the solar gravitational redshift. (Dravins 2008)}
\label{Fig2}
\end{figure}

\subsection{Spatially resolved stellar spectroscopy}

A major milestone in stellar astronomy will be the observation of stars not as point sources, but as extended surface objects.  Already, disks of some large stars can be imaged with large telescopes and interferometers, with many more resolvable with future facilities.

3-dimensional models predict spectral changes across stellar disks, differing among various stars and indicating their level of surface `corrugation' (Dravins \& Nordlund 1990a).  In stars with `smooth' surfaces, convective blueshifts decrease from disk center towards the limb, since vertical velocities become perpendicular to the line of sight, and the horizontal ones contributing Doppler shifts appear symmetric.  Stars with surface `hills' and `valleys' show the opposite, a blueshift increasing towards the limb, where one sees approaching (blueshifted) velocities on the slopes of those `hills' that are facing the observer.  Further effects appear in time variability: on a `smooth' star, temporal fluctuations are caused by the random evolution of granules. Near the limb of a `corrugated' star, further variability is added since the swaying stellar surface sometimes hides some granules from direct view.

To realize corresponding observations requires spectrometers with integral-field units and adaptive optics on extremely large telescopes or interferometers.  Options for two-dimensional imaging include Fourier transform spectrometers.  These provide high spectral resolution for extended objects (e.g., Maillard 2005), although their noise properties have hitherto limited their use in point-source observations

\section{Intergalactic wavelength shifts in quasars}

Wavelength shifts analogous to those in convective stellar photospheres may be expected also in intergalactic absorption lines from various metals, as seen in quasar spectra.  These lines typically form in the intergalactic medium of galaxy clusters that happen to fall in the line of sight towards the more distant quasar.  Such intergalactic gas undergoes non-random convective motions qualitatively analogous to those in stellar photospheres (even if the dynamic timescales, now on order 100 million years, are \textit{very} much longer).  Active galactic nuclei are normally located near such cluster cores and are sources of energetic cosmic rays, heating the surrounding plasma which becomes buoyantly unstable, thus driving convective motions within clusters of galaxies (e.g., Chandran 2004, 2005; Reynolds et al.\ 2005; Sharma et al.\ 2009; Vernaleo \& Reynolds 2006). 

Corresponding 3-dimensional hydrodynamic models reproduce several features of the observed structures in the intergalactic gas, and should become possible to use as a model input for computing synthetic spectral-line profiles and wavelength shifts.  Differential shifts for lines formed within the same cluster, among those with different oscillator strength, ionization level, etc., could then diagnose intergalactic hydrodynamics, even for the distant Universe.  For example, lines formed in the deeper gravitational potential closer to the cluster cores would be expected to be slightly more gravitationally redshifted than lines formed further out.  Lines of high excitation potential may be expected to predominantly form in hotter gas (where more atoms are in high-excitation states).  If these lines are strong, they will predominantly form in the near-side of the clusters, and therefore be seen as more blueshifted (the convective flow of hot gas is mainly radially outward in the cluster, thus approaching and blueshifted on its near-side, while the observable contributions from the receding and redshifted flow on the far side are weakened due to self-absorption in any stronger line).  Such shifts must be modeled and segregated from other possible shifts -- perhaps due to cosmologically changing `constants' -- if such other shifts are to be reliably deduced.

Even if such lineshift modeling is not yet available, one can try to estimate plausible orders of magnitude.  If the analogy to stellar line formation holds, one could expect wavelength shifts due to the different statistical weighting of line contributions from different inhomogeneities on the order of perhaps 1\% of the `turbulent' line broadening.  Since predicted gas-flow velocities are on order 50--100 km~s$^{-1}$, this implies lineshifts on order 500--1000 m~s$^{-1}$.  In order to segregate such intrinsic shifts in complex spectra, to map depth structure from multiple line components or perhaps even to map lateral structure from secular time changes, one clearly will need resolving powers of at least 300,000, and ideally even 1,000,000.  Further, since the sources are faint while low-noise spectra are required, extremely large telescopes are probably called for.

\begin{figure*}[t!]\centering
\includegraphics[width=15cm]{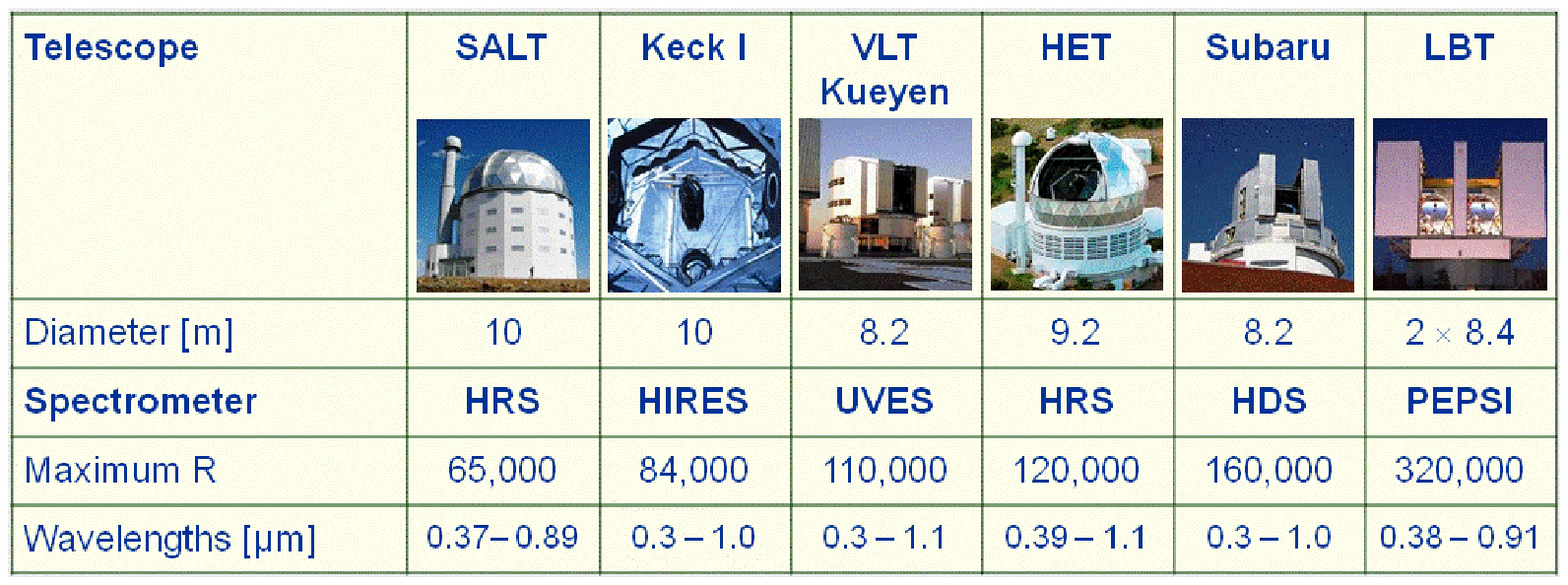} 
\caption{ Existing or planned high-resolution spectrometers for the visual range at existing 8-10 m class telescopes. }
\label{Fig3}
\end{figure*}

\section{Spectrometers at the largest telescopes} 

Existing or planned high-resolution stellar spectrometers for the visual at current very large telescopes are summarized in Fig.~4.  While some specialized instruments (R~$\approx$~10$^{6}$) have been used in particular to measure interstellar lines (e.g., the Ultra-High-Resolution Facility at the Anglo-Australian Telescope; Diego et al.\ 1995), the currently only seriously high-resolution night-time instrument foreseen at any very large telescope is PEPSI (\textit{Potsdam Echelle Polarimetric and Spectroscopic Instrument}) for the Large Binocular Telescope, planned to reach R~=~320,000 (Strassmeier et al.\ 2003).  Among the PEPSI science cases (Strassmeier et al.\ 2004), diagnosing 3-dimensional stellar hydrodynamics was examined, noting that studies of bisector curvature indeed require such resolutions.  Steffen \& Strassmeier (2007) further discuss the PEPSI `deep spectrum' project for the highest quality spectra for any star other than the Sun.

At the largest telescopes, challenges in highest-resolution spectroscopy stem from the difficulty to optically match realistically-sized grating spectrometers to the large image scales, i.e., to squeeze starlight into a narrow spectrometer entrance slit while avoiding unrealistically large optical elements (Span{\`o} et al.\ 2006).  The use of image slicers would limit the number of measurable spectral orders although some remedy could be offered by adaptive optics (Ge et al.\ 2002; Sacco et al.\ 2004).

Pasquini et al.\ (2005) and D'Odorico et al.\ (2007) discuss future spectrometers.  High efficiency requires large optics: the two-grating mosaic for PEPSI makes up a 20$\times$80~cm R4 echelle, while ESPRESSO for the VLT combined focus, and CODEX for the E-ELT aim at twice that: 20$\times$160 cm (Pasquini et al.\ 2008; Span{\`o} et al.\ 2008).  Nevertheless, resolutions would not reach much above R~=~100,000.

\subsection{Avoiding telluric absorption?}

Even with superlative spectral resolution and photometric precision, spectra are distorted by telluric absorption and emission lines and bands, compromising the spectral fidelity.  Actually, in some wavelength regions, the level of contamination makes it doubtful whether it will ever be possible to record truly high-fidelity solar or stellar spectra from the ground.

In some regions, telluric lines from, say, H$_{2}$O, are sharp and well confined in wavelength, and then (at least in principle) can be identified and removed in the analysis (e.g., Hadrava 2006).  However, an accurate removal requires the terrestrial lines to be spectrally resolved, and not even the `high' resolution in solar atlases is fully adequate here.  Inspection of telluric lines in clean spectral regions of the Kitt Peak Atlas of integrated sunlight (Kurucz et al.\ 1984) shows that these lines are still unresolved (for an example, see $\lambda$~1250.7 nm), instead displaying the characteristic `ringing' instrumental profile of a Fourier transform spectrometer, even at the resolution R~=~500,000.

There are other and more treacherous telluric features, extending over wider wavelength regions.  These include diffuse absorptions due to ozone O$_{3}$ and the oxygen dimer [O$_{2}$]$_{2}$ that appear throughout the visual. Ozone produces the atmospheric transmission cutoff in the near ultraviolet, where old stellar spectrograms can actually be used to infer past amounts of terrestrial ozone (Griffin 2005).

Only in a few cases has it been possible to compare high-resolution spectra recorded from above and beneath the atmosphere.  Sirius was observed with the Goddard High-Resolution Spectrograph on the Hubble Space Telescope in the ultraviolet up to 320 nm, while ground-based observations reach down to 305 nm.  The overlapping interval is very instructive in demonstrating how the apparent continuum in ground-based recordings can locally be wrong by some 10\% due to diffuse telluric absorptions (Griffin 2005, and private comm.).

Kurucz (2005) computed synthetic telluric spectra, including O$_{3}$ and [O$_{2}$]$_{2}$, showing how they affect the highest-resolution solar spectra.  The richness of the telluric spectra means that weak lines with depths of perhaps only some percent or less are often superposed onto the flanks or cores of solar ones, making the reduction for their effects awkward or practically impossible (especially since atmospheric extinction is variable in time; Stubbs et al.\ 2007).  In fact, for many spectral regions, this sets a signal-to-noise limit on the order of 100 or worse, irrespective of the photometric precision reached.  The affected regions can be inspected in a spectral atlas of telluric absorption (R~$\approx$~300,000), derived from solar data by Hinkle et al.\ (2003).

Observations from orbit would eliminate telluric absorptions (although geocoronal emission would still be present), but any space mission for high-resolution stellar spectroscopy would probably be complex, considering not only that a large telescope is needed but also that the Doppler shift induced by the spacecraft motion implies continuous wavelength changes.

\subsection{Accurate wavelength calibration?}

Grating and Fourier transform instruments establish their wavelength scales through different schemes, and each is likely to have its characteristic signatures in its wavelength noise.

Grating spectrometers commonly use an emission-line lamp (thorium, etc.), and thus their wavelength scale depends upon the accuracy of the corresponding laboratory wavelengths and on how well the lamp at the telescope reproduces the laboratory sources.  Traditionally, the wavelength noise among individual lines may have been around 100~$\mathrm{m\, s^{-1}}$ but recent calibrations of thorium-argon hollow-cathode lamps have now decreased the internal scatter to some 10 $\mathrm{m\, s^{-1}}$ (Lovis \& Pepe 2007), further permitting a selection of subsets of lines for best calibration (Murphy et al.\ 2007a).  

Fourier transform spectrometers have their specific issues.  The interferogram is recorded sequentially for its sinusoidal components, between low Fourier frequencies and high ones (how high determines the spectral resolution).  However, the realities of detectors and photon noise normally preclude the entire spectrum being measured at once; instead only one piece at a time must be selected by some pre-filter or similar device.  The transmission functions of these pre-filters have to be precisely calibrated to enable a correct continuum level to be set.  Its placement to better than 1:100 is no trivial task, as illustrated by the efforts to improve the continuum in the Kitt Peak solar atlases by Neckel (1999) and Kurucz (2005). 

\subsection{Future high-fidelity spectroscopy?}

Although several parameters limiting highest-fidelity spectroscopy can be identified, their immediate remedies cannot.  Avoiding the terrestrial atmosphere requires ambitious space missions, and improving laboratory wavelengths requires dedicated long-term efforts.  Possibly, the least difficult part is to improve instrumental calibrations.

Laboratory devices, in particular the laser frequency comb, permit remarkably precise wavelength determinations, and are also envisioned for future astronomical spectrometers (Araujo-Hauck et al.\ 2007; Murphy et al.\ 2007b; Steinmetz et al.\ 2008).  However, it is not necessarily the stability of the calibration device that limits the accuracy in astrophysical spectra because it may in addition require identical light paths for the source and its calibration, and without the calibration light contaminating the source spectrum.

The telescope-spectrometer interface is another issue.  Analyses of the UVES spectrometer during asteroid observations reveal a noise of typically 10--50~m~s$^{-1}$, apparently caused by a non-uniform and variable illumination in the image projected by the telescope onto the spectrometer entrance slit (Molaro et al.\ 2008).  Given the chromatic nature of atmospheric dispersion, the wavelength dependence of such shifts could also mimic line-depth dependences since stronger atomic lines often occur at shorter wavelengths.  Although a more uniform illumination is provided by image slicers or fiber-optics feeds, their use becomes awkward if one requires both extended spectral coverage and the use of (very) large telescopes.  For non-solar work, the former requires multi-order echelle spectrometers, and the latter implies (very) large image scales.  Light from all image slices, projected onto the focal plane, would overlap adjacent echelle orders (precluding the recording of extended spectral regions), and an optical fiber would need to have a large diameter to embrace most starlight, requiring entrance apertures that are too wide for adequate spectral resolution (or else cause severe light losses).  As already mentioned, possible solutions that avoid the construction of huge instruments could include spectrometers with adaptive optics.

Some instrumentation planning committees appear to concentrate on maximum light efficiency for future instruments.  While of course desirable, it is a fallacy to believe that highest optical efficiency would be crucial to scientific discovery: what is required is {\textit{adequate}} efficiency to reveal novel features in astronomical sources.  Still, a grand challenge remains in designing an efficient truly high-resolution (R~$\approx$~10$^{6}$) and high-fidelity spectrometer for future extremely large telescopes!

\
\acknowledgements
This work is supported by The Swedish Research Council and The Royal Physiographic Society in Lund. It used data from the UVES Paranal Observatory Project of the European Southern Observatory (ESO DDT Program ID 266.D-5655).  It also used solar spectral atlases obtained with the Fourier Transform Spectrometer at the McMath/Pierce Solar Telescope situated on Kitt Peak, Arizona, operated by the National Solar Observatory, a Division of the National Optical Astronomy Observatories.


\end{document}